\documentclass[%
 reprint,
 superscriptaddress,
 amsmath,amssymb,
 aps,
 pra,
floatfix,
]{revtex4-1}

\usepackage{graphicx}% Include figure files
\usepackage{dcolumn}% Align table columns on decimal point
\usepackage{bm}% bold math
\usepackage{times}
\usepackage{hyperref}% add hypertext capabilities
\usepackage{amssymb}

\usepackage{amsmath}
\usepackage{algorithm}
\usepackage{algpseudocode}
\usepackage{mathtools}

\usepackage{tikz}
\usetikzlibrary{positioning}
\definecolor {processblue}{cmyk}{0.96,0,0,0}
\definecolor {processyellow}{cmyk}{0,0,0.18,0}

\makeatletter
\def\BState{\State\hskip-\ALG@thistlm}
\makeatother

\makeatletter
\newcommand*\bigcdot{\mathpalette\bigcdot@{.6}}
\newcommand*\bigcdot@[2]{\mathbin{\vcenter{\hbox{\scalebox{#2}{$\m@th#1\bullet$}}}}}
\makeatother

\algnewcommand{\IIf}[1]{\State\algorithmicif\ #1\ \algorithmicthen}
\algnewcommand{\EndIIf}{\unskip\ \algorithmicend\ \algorithmicif}
\DeclareMathOperator{\rre}{Re}
\DeclareMathOperator{\iim}{Im}
\DeclareMathOperator{\tr}{tr}

\begin{document}

%\title{Quantum optimal control with automatic differentiation using graphics processors} 
\title{Speedup for quantum optimal control from automatic differentiation\protect\\ based on graphics processing units}

\author{Nelson Leung}
 \email{nelsonleung@uchicago.edu}
\affiliation{%
 The James Franck Institute and Department of Physics, University of Chicago, Chicago, Illinois 60637, USA 
}%
\author{Mohamed Abdelhafez}%

\affiliation{%
 The James Franck Institute and Department of Physics, University of Chicago, Chicago, Illinois 60637, USA 
}%

\author{Jens Koch}%
\affiliation{%
Department of Physics and Astronomy, Northwestern University, Evanston, Illinois 60208, USA 
}%

\author{David Schuster}%
\affiliation{%
 The James Franck Institute and Department of Physics, University of Chicago, Chicago, Illinois 60637, USA 
}%

\date{\today}% It is always \today, today,

\begin{abstract}
We implement a quantum optimal control algorithm based on automatic differentiation and harness the acceleration afforded by graphics processing units (GPUs). Automatic differentiation allows us to specify advanced optimization criteria and incorporate them in the optimization process with ease. We show that the use of GPUs can speed up calculations by more than an order of magnitude. Our strategy facilitates efficient numerical simulations on affordable desktop computers, and exploration of a host of optimization constraints and system parameters relevant to real-life experiments. We demonstrate optimization of quantum evolution based on fine-grained evaluation of performance at each intermediate time step, thus enabling more intricate control on the evolution path, suppression of departures from the truncated model subspace, as well as minimization of the physical time needed to perform high-fidelity state preparation and unitary gates.
\end{abstract}

\pacs{Valid PACS appear here}% PACS, the Physics and Astronomy
\maketitle

\section{Introduction}

The techniques and algorithms used to optimize the control of quantum systems 
\cite{Glaser2015Training,Khaneja2005Optimal,deFouquieres2011Second,Sklarz2002Loading,Eitan2011Optimal,Gollub2008Monotonic,Nigmatullin2009Implementation,Reich2012Monotonically,Palao2003Optimal,Tannor1992Control,Zhu1998Rapidly,Zhu1998Rapid,Ohtsuki2004Generalized,Maday2003New,Ohtsuki2007Monotonically,Borzi2008Formulation,Ditz2008Cascadic} and those underlying the field of deep neural networks \cite{Haykin1994Neural,HechtNielsen1989Theory} share a number of common elements. Both areas heavily use linear algebra operations combined with gradient descent optimization.
Thus, advanced hardware and software technology recently emerging from the rapid development of machine learning also paves the way for a significant boost of optimal quantum control techniques. 

A crucial factor for recent impressive progress in machine learning has been the leveraging of massive parallelism native to graphics processing units (GPUs) \cite{Oh2004GPU,Catanzaro2008Fast,Sharp2008Implementing,Raina2009Largescale,Steinkraus2005Using}. Similarly GPUs have been used to accelerate computations in many areas of quantum physics and chemistry \cite{Block2010MultiGPU,Clark2010Solving,Wilkinson2011Acceleration,ufimtsev2008graphical,vogt2008accelerating,olivares2009accelerating,Hou2016Full}.
Specifically, GPUs are extremely efficient in multiplying very large matrices \cite{Cui2009Improving,Fatahalian2004Understanding}. Such multiplications also form a central step in the simulation and optimal control of quantum systems. Exploiting this advantageous feature of GPUs, we achieve significant speed improvements in optimizing control schemes for systems at the current frontiers of experimental quantum computation. As the number of qubits in these experiments is increasing \cite{Corcoles2015Demonstration,Debnath2016Demonstration,Kelly2015State}, it becomes increasingly important to take advantage of optimal control techniques. Moreover, recent advances in commercially available electronics -- e.g., arbitrary waveform generators enabling base-band synthesis of the entire microwave spectrum \cite{M8196A} -- afford new capabilities which quantum optimal control is uniquely well-suited to harness.

There have been numerous theoretical developments of numerical and analytical methods for quantum optimal control (see Ref.\ \onlinecite{Glaser2015Training} for a recent review). The algorithms involved are predominantly based on gradient methods, such as realized in gradient ascent pulse engineering (GRAPE) \cite{Khaneja2005Optimal,deFouquieres2011Second}, Krotov algorithms \cite{Sklarz2002Loading,Eitan2011Optimal,Gollub2008Monotonic,Nigmatullin2009Implementation,Reich2012Monotonically,Palao2003Optimal,Tannor1992Control} or rapid monotonically convergent algorithms \cite{Zhu1998Rapidly,Zhu1998Rapid,Ohtsuki2004Generalized,Maday2003New,Ohtsuki2007Monotonically,Borzi2008Formulation,Ditz2008Cascadic}, and are available in several open-source packages, including QuTiP \cite{Johansson2012QuTiP,Johansson2013QuTiP}, DYNAMO \cite{Machnes2011Comparing}, Spinach \cite{Hogben2011Spinach}, and SIMPSON \cite{Tosner2009Optimal}. Quantum optimal control has been remarkably successful in determining optimized pulse sequences \cite{Borneman2010Application}, designing high-fidelity quantum gates \cite{Sporl2007Optimal,Rebentrost2009Optimal,Nigmatullin2009Implementation,Nebendahl2009Optimal,Kelly2014Optimal,Machnes2015Gradient,Egger2014Optimized,Liebermann2016Optimal,Rebentrost2009Optimal_2,Egger2014Adaptive,Kosut2013Robust}, and preparing entangled states \cite{Dolde2014Highfidelity,Platzer2010Optimal,Watts2015Optimizing,Goerz2015Optimizing,Goerz2016Charting}. 

Optimal control is a versatile concept which can be applied to a vast variety of quantum systems. Typically there is a primary goal (e.g. maximizing fidelity to a target state/unitary), as well as additional constraints/costs associated with specific experimental systems. Examples of such constraints include fixed maximum amplitudes of control pulses \cite{Skinner2004Reducing,Kobzar2004Exploring}, maximum overall power of control signals \cite{Kobzar2008Exploring}, and limited time resolution of arbitrary waveform generators \cite{Motzoi2011Optimal}. Further, finite coherence of quantum systems motivates minimizing the overall time needed for reaching the intended state or unitary (time-optimal control) \cite{Chen2015Neartimeoptimal}. In certain cases, steering the quantum system among an optimal path (time-dependent target) may be desired \cite{Serban2005Optimal}. Incorporating new constraints in the optimization process often requires the analytical derivation and implementation of additional contributions to the gradient calculation, and may necessitate significant effort to deploy on large computer clusters. This issue can greatly impede the ability to quickly develop control strategies for new problems.

To overcome these obstacles, we have implemented a quantum optimal control scheme that incorporates constraints via automatic differentiation \cite{BartholomewBiggs2000Automatic,Wengert1964Simple} and utilizes GPUs for boosting computational efficiency. Specifically, automatic differentiation handles the updating of gradient calculations in the backward propagation algorithm \cite{HechtNielsen1989Theory}, and thus eliminates the need to hard-code additional gradient contributions from constraints. For the actual optimal control applications we present in this paper, we find that the computational speed-up from utilizing GPUs becomes significant for Hilbert space sizes exceeding dimensions of the order of one hundred, see Fig.\ \ref{fig:benchmark}. Together, these features allow a quick turnaround for varying optimization constraints and system parameters, rendering this approach invaluable for the study of quantum optimal control. In this paper, we describe the implementation of automatic differentiation, demonstrate its application to quantum optimal control of example systems relevant to quantum computing and quantum optics, and discuss the performance gains achieved by utilizing GPUs. 

\section{Theory}
We briefly review the essential idea of quantum optimal control and introduce the notation used throughout our paper. We consider the general setting of a quantum system with intrinsic Hamiltonian $\mathcal{H}_0$ and a set of external control fields $\{u_1(t),\ldots,u_M(t)\}$ acting on the system via control operators $\{\mathcal{H}_1,\ldots,\mathcal{H}_M\}$. The resulting system Hamiltonian is given by $H(t)=\mathcal{H}_0 + \sum_{k=1}^M u_k(t) \mathcal{H}_k$. Optimal control theory aims to minimize deviations from a target state or target unitary by appropriate adjustments of the control fields $u_k(t)$. To implement this optimization, the time interval of interest is discretized into a large number $N$ of sufficiently small time steps $\delta t$. Denoting intermediate times by $t_j = t_0 + j\,\delta t$, the Hamiltonian at time $t_j$ takes on the form
\begin{equation}
H_j = \mathcal{H}_0 + \sum_{k=1}^M u_{k,j}\mathcal{H}_{k}.
\label{eq:Hamiltonian}
\end{equation}
The control fields subject to optimization now form a set $\{u_{k,j}\}$ of $d=M\cdot N$ real numbers.
 
The quantum evolution from the initial time $t=t_0$ to time $t_j$ is described by a propagator $K_j$, decomposed according to
\begin{equation}
K_j = U_j U_{j-1} U_{j-2}\dots U_{1}U_{0}
\label{eq:evolution}
\end{equation}
where 
\begin{equation}
U_j = \exp(-i H_j \delta t)
\label{eq:propagator}
\end{equation}
is the propagator for the short time interval $[t_j,t_j+\delta t]$. (Here and in the following, we set $\hbar=1$.) 
Evolution of a select initial state $|\Psi_{0}\rangle$ from $t = t_0$ to $t = t_j$ then takes the usual form
\begin{equation}
|\Psi_{j}\rangle = K_j|\Psi_{0}\rangle.
\label{eq:wave_function}
\end{equation}
In the decomposition of $K_j$, each short-time propagator $U_i$ can be evaluated exactly by matrix exponentiation or approximated by an appropriate series expansion. Propagation methods which go beyond the piecewise-constant approximation for the propagation, can further improve speed and accuracy \cite{Machnes2015Gradient}.

Optimization of the discretized control fields $\textbf{u} \in \mathbb{R}^d$ can be formulated as the minimization of a cost function $C(\textbf{u})$ where $C:\,\mathbb{R}^d\to\mathbb{R}^+$. 
%This $\textbf{u}$ is a matrix whose every row corresponds to one control operator and its columns correspond to the different time steps.
Table \ref{table:cost_functions} shows some of the most important cost function contributions used for quantum optimal control. The total cost function is a linear combination of these cost functions, $C = \sum_\mu \alpha_{\mu} C_\mu$. The weight factors $\alpha_\mu$ must be determined empirically, and depend on the specific problem and experimental realization at hand. 
In the following, we discuss these relevant cost function contributions.

\subsection{Important types of cost function contributions\label{sec:costfunctions}}
The first cost contribution, $C_1(\mathbf{u})$, is the primary tool for realizing a target unitary $K_T$, such as a single or multi-qubit gate. Cost is incurred for deviations between the target unitary and the realized unitary $K_N$ at a given final time $t_N$. For a system with Hilbert space dimension $D$, its expression $1 - |\tr(K_T^\dagger K_N)/D|^2$ \cite{Khaneja2005Optimal} represents the infidelity obtained from the trace distance between the target unitary and the realized unitary. Minimizing this cost function is the principle goal of the quantum control problem.

The second cost function, $C_2(\mathbf{u})=1- |\langle\Psi_T|\Psi_{N}\rangle|^2$ measures the distance between a desired target state $|\Psi_T\rangle$ and the state $|\Psi_N\rangle$ realized at the final time $t_N$, as obtained from evolution of a given initial state $|\Psi_0\rangle$. In addition, generalizing $C_2$ to multiple initial and target states
is useful for performing a unitary $K_T$ which is only defined on some subspace $H_\mathcal{S}$ of the modeled Hilbert space. Such restriction to a selected subspace is of practical importance whenever a desired unitary is to be implemented within some computational subspace only, as is common for quantum computation applications. There, evolution of higher excited states or auxiliary systems outside the computational subspace is immaterial. Optimal control, then, can be achieved by simultaneous evolution of a set of initial states $\{|\Psi_{0}^s\rangle\}$ ($s=1,2,\ldots,S$) that forms a basis of $H_\mathcal{S}$. Optimal control fields are obtained from minimizing the composite state infidelity $C_{2\Sigma}(\mathbf{u})=1-|\frac{1}{S}\sum_s\langle\Psi_T^s|P_\mathcal{S}|\Psi_{N}^s\rangle|^2$ relative to the desired target states $|\Psi_T^s\rangle = K_T|\Psi_{0}^s\rangle$. (Here, $P_\mathcal{S}$ is the projector onto subspace $H_\mathcal{S}$.) 

This composite state-transfer cost function when used over a complete basis is equivalent to the gate fidelity, but has several advantages.  Most importantly it is more memory efficient requiring only the current state to be stored rather than the whole unitary. In addition, it is very amenable to distributed computing approaches. However, when the unitary transfer matrix can be stored in memory, propagating the full unitary can take advantage of the parallelism of the GPU for smaller problems (see Fig.~\ref{fig:benchmark}).

%If a complete basis of initial states is used, usage of $C_2$ becomes formally equivalent to implementation of $C_1$. However, using state optimization with $C_2$ can be advantageous when encountering memory bottlenecks, since this strategy only requires storage of quantum states, rather than operators, and is hence significantly more memory efficient.

\begin{table}
\begin{ruledtabular}
\begin{tabular}{lcc}
$\mu$ &\textbf{Cost function contribution} & $C_\mu(\textbf{u})$ \\ \hline 
$1$ & Target gate infidelity & $ 1- |\tr(K_T^\dagger K_N)/D|^2$ \\ 
$2$ & Target state infidelity & $ 1- |\langle\Psi_T|\Psi_{N}\rangle|^2$  \\ 
$3$ & Control amplitudes & $|\textbf{u}|^2$  \\ 
$4$ & Control variations &$\sum_{j,k} |u_{k,j} - u_{k,j-1}|^2$  \\ 
$5$ & Occupation of forbidden state & $\sum_{j} |\langle \Psi_F| \Psi_{j} \rangle|^2$ \\
$6$ & Evolution time (target gate) & $ 1- \frac{1}{N}\sum_{j} |\tr( K_T^\dagger K_j )/D|^2$ \\
$7$ & Evolution time (target state) & $ 1- \frac{1}{N}\sum_{j} |\langle \Psi_T| \Psi_{j}\rangle|^2 $ \\ 
\end{tabular}
\caption{Relevant contributions to cost functions for quantum optimal control. Names of contributions indicate the quantity to be \emph{minimized}.}
\label{table:cost_functions}
\end{ruledtabular}
\end{table}

Like many optimization problems, quantum optimal control is typically underconstrained. In order to obtain control fields that are consistent with specific experimental capabilities and limitations, it is often crucial to add further constraints on the optimization. Control fields must be realizable in the lab, should be robust to noise, and avoid large control amplitudes and rapid variations based on signal output specifications of instruments employed in experiments. Exceedingly strong control fields may also be problematic due to heat dissipation which may, for instance, raise the temperature inside a dilution refrigerator. These points motivate the consideration of additional cost function contributions in the following.

One such contribution, $C_3(\mathbf{u})=|\mathbf{u}|^2$ suppresses large control-field amplitudes globally, and is commonly employed in quantum optimal control studies \cite{Khaneja2005Optimal,Skinner2004Reducing,Kobzar2004Exploring,Heeres2016Implementing}. (The generalization to more fine-grained suppression of individual control fields is straightforward to implement as well.) Penalizing the $L^2$ norm of the control fields favors solutions with low amplitudes. It also tends to spread relevant control fields over the entire allowed time window. While $C_3$ constitutes a ``soft'' penalty on control-field amplitudes, one may also apply a trigonometric mapping to the amplitudes to effect a hard constraint strictly enforcing fixed maximum amplitudes \cite{Farnum2005Trigonometric}.

The fourth type of contribution to the cost function, $C_4(\mathbf{u})=\sum_{j,k} |u_{k,j} - u_{k,j-1}|^2$, penalizes rapid variations of control fields by suppressing their (discretized) time derivatives \cite{Heeres2016Implementing}. The resulting smoothening of signals is of paramount practical importance, since any instrument generating a control field has a finite impulse response. If needed, contributions analogous to $C_4$ which suppress higher derivatives or other aspects of the time dependence of fields can be constructed. Together, limiting the control amplitudes and their time variation filters out high-frequency ``noise'' from control fields, which is an otherwise common result of less-constrained optimization. Smoother control fields also have the advantage that essential control patterns can potentially be recognized and given a meaningful interpretation.

The contribution $C_5(\mathbf{u})=\sum_{j} |\langle \Psi_F| \Psi_{j} \rangle|^2$ to the cost function has the effect of suppressing occupation of a select ``forbidden'' state $|\Psi_F\rangle$ (or a set of such states, upon summation) throughout the evolution. The inclusion of this contribution addresses an important issue ubiquitous for systems with Hilbert spaces of large or infinite dimension. In this situation, truncation of Hilbert space is needed or inevitable due to computer memory limitations. (Note that this need even arises for a single harmonic oscillator.) Whenever the evolution generated by optimal control algorithms explores highly excited states, truncation introduces a false non-linearity which can misguide the optimization. Including additional states can, in principle, mitigate this problem, but is generally computationally very expensive. An independent physics motivation for avoiding occupation of highly-excited states consists of spontaneous relaxation in realistic systems: high-energy states are often more lossy (as is usually the case, e.g., for superconducting qubits), and possibly more difficult to model. Active penalization of such states therefore has the two-fold benefit of keeping Hilbert space size at bay, and reducing unwanted fidelity loss from increased relaxation. To address these challenges, we employ an intermediate-time cost function \cite{Serban2005Optimal,Palao2008Protecting}: the cost function $C_5$ limits leakage to higher states during the entire evolution, and at the same time prevents optimization to be misinformed by artificial non-linearity due to truncation. We note that the efficacy of this strategy is system dependent: it works well, for example, for harmonic oscillators or transmon qubits \cite{Koch2007Chargeinsensitive} which have strong selection rules against direct transitions to more distant states, but may be less effective in systems such as the fluxonium circuit \cite{Manucharyan2009Fluxonium} where low-lying states have direct matrix elements to many higher states.

\begin{figure}
  \centering \includegraphics[width=1.0\columnwidth]{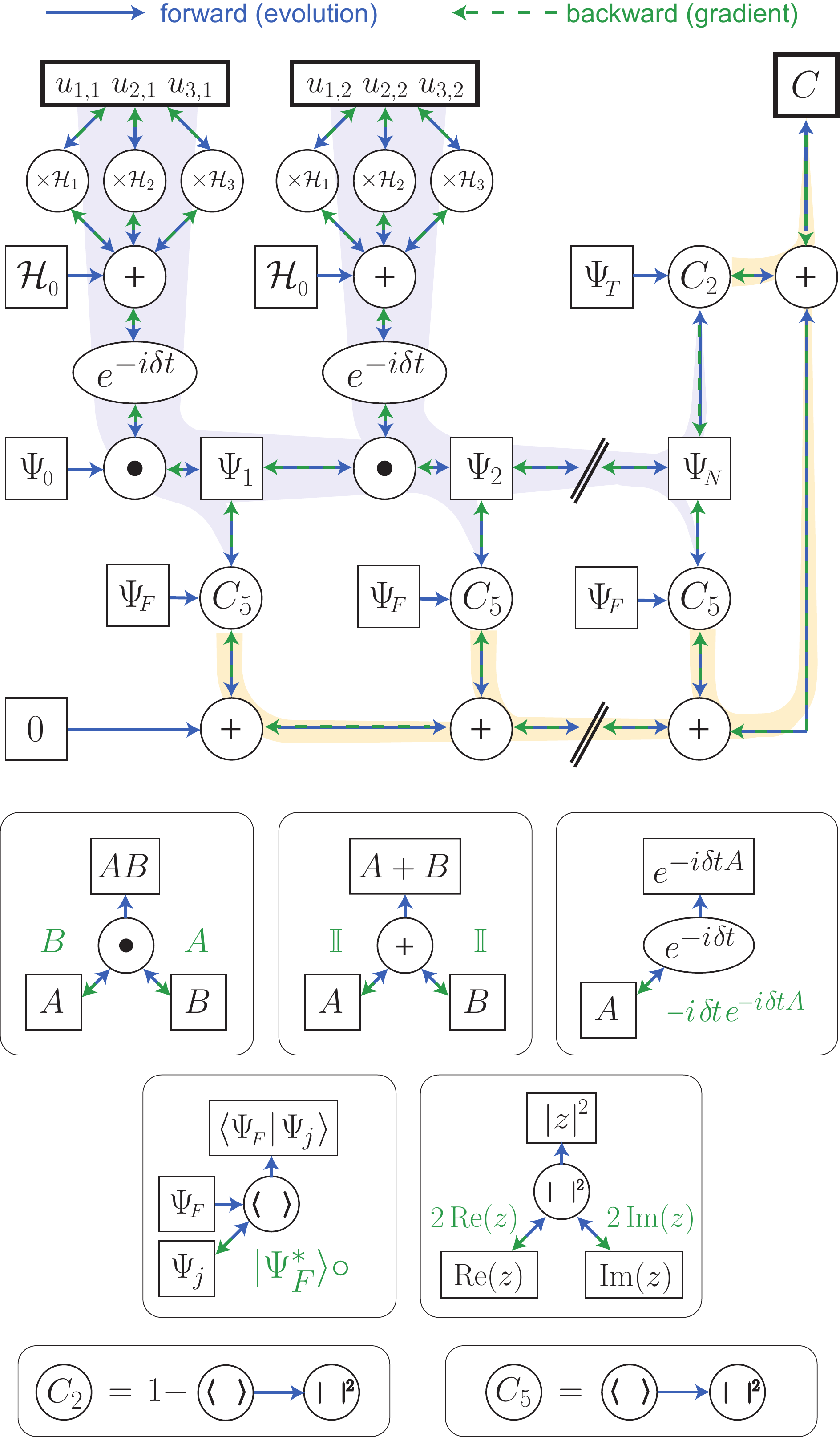}
  \caption{{Computational network graph for quantum optimal control. } Circular nodes in the graph depict elementary operations with known derivatives (matrix multiplication, addition, matrix exponential, trace, inner product, and squared absolute value). Backward propagation for matrices proceeds by  matrix multiplication, or where specified, by the Hadamard product $\circ$. In the forward direction, starting from a set of control parameters $u_{k,j}$, the computational graph effects time evolution of a quantum state or unitary, and the simultaneous computation of the cost function $C$. The subsequent ``backward propagation'' extracts the gradient $\nabla_\mathbf{u}C(\mathbf{u})$ with respect to all control fields by reverse-mode automatic differentiation. This algorithm is directly supported by TensorFlow \cite{Abadi2016TensorFlow}, once such a computational network is specified.}
  \label{figure:nn_graph}
\end{figure}

Customarily, algorithms minimizing the cost function $C=\sum_\mu \alpha_\mu C_\mu$ for a given evolution time interval $[t_0,t_N]$ aim to match the desired target unitary or target state at the very end of this time interval. To avoid detrimental effects from decoherence processes during the evolution, it is often beneficial to additionally minimize the gate duration (or state preparation) time $\Delta t = t_N - t_0$ itself. Instead of running the algorithms multiple times for a set of different $\Delta t$, we employ cost function contributions of the form $C_6(\mathbf{u})=1- \frac{1}{N}\sum_{j} |\tr( K_T^\dagger K_j )/D|^2$ for a target unitary, or $C_7(\mathbf{u})=1- \frac{1}{N}\sum_{j} |\langle \Psi_T| \Psi_{j}\rangle|^2$ for a target state, respectively. These expressions penalize deviations from the target gate or target state not only at the final time $t_N$, but \emph{at every time step}. This contribution to the overall cost function therefore guides the evolution towards a desired unitary or state in as short a time as possible under the conditions set by the other constraints, and thus results in a time-optimal gate.

 We will demonstrate the utility of these cost function contributions in the context of quantum information processing in Section \ref{sec:examples}. The versatility of automatic differentiation allows straightforward extension to other contexts such as optimization of quantum observables.
 
\subsection{Gradient evaluation}
The weighted sum of cost functions, $C=\sum_\mu \alpha_\mu C_\mu$, can be minimized through a variety of gradient-based algorithms. Such algorithms are a very popular means of optimization thanks to their good performance and effectiveness in finding optimized solutions for a wide range of problems. At the most basic level, gradient-based algorithms minimize the cost function $C(\textbf{u})$ by the method of steepest descent, updating the controls $\mathbf{u}$ in the opposite direction of the local cost-function gradient $\nabla_\textbf{u} C(\textbf{u})$:
\begin{equation}
\textbf{u}' = \textbf{u} - \eta\, \nabla_\textbf{u} C(\textbf{u}) .
\label{eq:gradient}
\end{equation}
The choice of the update step size $\eta$ for the control field parameters $\mathbf{u}$, plays an important role for the convergence properties of the algorithm. A number of schemes exist which adaptively determine an appropriate step size $\eta$ in each iteration of the minimization algorithm. Our implementation supports second order methods such as L-BFGS-B \cite{Byrd1995Limited} as well as gradient descent methods developed for machine learning such as ADAM \cite{Kingma2015Adam}.

% A straightforward but naive way to obtain these derivatives is by perturbing each control parameter $\textbf{u}$ and repeatedly simulating the unitary evolution to get the cost function corresponding to the perturbed controls. However, this approach would result in computation steps of order $O(N^2)$ per iteration, where $N$ is the number of time steps. 

\begin{figure}

\begin {center}
\begin {tikzpicture}[
	-latex,
	auto,
	node distance =1.8 cm and 2.8cm,
	on grid,
	semithick,
	state/.style ={circle,
	                top color =white, bottom color = processblue!20,
		            draw, processblue, text=blue , minimum width =0.8cm},
	start/.style = {draw, 
	                top color =white, bottom color = processyellow,
	                align=center, anchor=west, minimum height=0.8cm, minimum width=0.8cm,inner sep=5},
	scale=0.7, every node/.style={scale=0.7}
	]
\node (A) at (-1.5,0)  [start] {$u_1$};
\node (B) at (4.5,0)  [start] {$u_2$};
\node (C) at (3.7,1.4)    [state] {$\sqrt{{\cdot}}$};
\node (D) at (2.4,2.1)  [state] {$\bullet$};
\node (E) at (0.2,2.1) [state] {$\sin$};
\node (F) at (2,3.4) [state] {$+$};
\node (G) at (1.5,4.7) [start] {$C(\mathbf{u})$};
\path (A) edge [bend right = 5] node[below =0.15 cm] {} (D);
\path (A) edge [bend right = -5]node[below =0.15 cm] {} (E);
\path (C) edge [bend right = 5]node[below =0.15 cm] {} (D);
\path (B) edge [bend right = 5] node[below =0.15 cm] {} (C);
\path (E) edge [bend right = -5] node[below =0.15 cm] {} (F);
\path (D) edge [bend right = 5] node[below =0.15 cm] {} (F);
\path (F) edge [bend right = 0] node[below =0.15 cm] {} (G);
\end{tikzpicture},
\end{center}

\caption{{Sample computational graph for automatic differentiation. } Automatic differentiation utilizes the decomposition of the multivariable cost function $C(\mathbf{u})$ into its computational graph of elementary operations, each of which has a known derivative. In reverse-accumulation mode, all partial derivatives of $C$ are evaluated in a recursion from the top level ($C$) back towards the outermost branches (variables $\mathbf{u}$). }
  \label{figure:sample_nn_graph}

\end{figure}
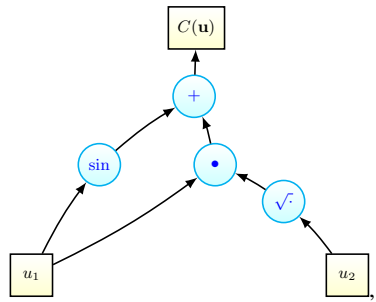

For the evaluation of the gradient $\nabla_\mathbf{u}C$ we make use of automatic differentiation \cite{Wengert1964Simple,BartholomewBiggs2000Automatic} in reverse-accumulation mode. In brief, this algorithm utilizes the decomposition of the multivariable cost function $C(\mathbf{u})$ into its computational graph of elementary operations (addition, matrix multiplications, trace, etc.), each of which has a known derivative. In reverse-accumulation mode, all partial derivatives of $C$ are evaluated in a recursion from the top level ($C$) back towards the outermost branches (variables $\mathbf{u}$) -- rather similar to the procedure of obtaining a derivative with pencil and paper. For instance, for the simple function
\begin{align}\nonumber%\label{eq:decomp}
&C(\mathbf{u}) = \sin(u_1) + u_1\cdot \sqrt{u_2} %\\\nonumber
%&\qquad\qquad
= f_+\bigg[\sin(u_1) ,\,\, f_{\bigcdot}(u_1,\sqrt{u_2}) \bigg]
\end{align}
one obtains all partial derivatives by a recursion starting with the evaluation of
\begin{equation}\nonumber
\frac{\partial}{\partial u_j}C =
D_1 f_+\,[\frac{\partial}{\partial u_j}\sin] \,+\, 
D_2 f_+\, [\frac{\partial}{\partial u_j}f_{\bigcdot}] = \cdots.
\end{equation}
Here, $D_j f$ stands for the derivative of a multivariable function $f$ with respect to its $j$-th argument; square brackets denote subsequent numerical evaluation of the enclosed term. (Function arguments are suppressed for brevity.)

Automatic differentiation has become a central tool in machine learning \cite{Baydin2015Automatic}, and equally applies to the problem of optimal control of quantum systems. In this approach, the gradient of a set of elementary operations is defined and more complex functions are built as a graph of these operations. The value of the function is computed by traversing the graph from inputs to the output, while the gradient is computed by traversing the graph in reverse via the gradients. This methodology gives the same numerical accuracy and stability of analytic gradients without requiring one to derive and implement analytical gradients specific to each new trial cost function.

All cost functions summarized in table \ref{table:cost_functions} can be conveniently expressed in terms of common linear-algebra operations. 
Figure \ref{figure:nn_graph} shows the network graph of operations in our software implementation, realizing quantum optimal control with reverse-mode automatic differentiation. For simplicity, the graph only shows the calculation of the cost functions $C_2$ and $C_5$. The cost function contributions $C_1, C_6$, and $C_7$ are treated in a similar manner. The suppression of large control amplitudes or rapid variations, achieved by $C_3$ and $C_4$, is simple to include, since the calculation of these cost function contributions is based on the control signals themselves and does not involve the time-evolved state or unitary. The host of steps for gradient evaluation is based on basic matrix operations like summation and multiplication.

Reverse-mode automatic differentiation \cite{HechtNielsen1989Theory} provides an efficient way to carry out time evolution and cost function evaluation by one forward sweep through the computational graph, and calculation of the full gradient by one backward sweep. In contrast to forward accumulation, each derivative is evaluated only once, thus enhancing computational efficiency. The idea of backward propagation is directly related to the GRAPE algorithm for quantum optimal control pioneered by Khaneja and co-workers \cite{Khaneja2005Optimal}, see Appendix \ref{app:gradient}. While the original GRAPE algorithm bases minimization exclusively on the fidelity of the final evolved unitary or state, advanced cost functions (such as $C_5$ through $C_7$) require the summation of cost contributions from each intermediate step during time evolution of the system. Such cost functions go beyond the usual GRAPE algorithm, but can be included in the more general backward propagation scheme described above. [Appendix \ref{app:gradient} shows analytical forms for gradients for cost functions that are based on time evolution ($\{C_1,C_2,C_5\}$).]

\section{Implementation}
Our quantum optimal control implementation utilizes the TensorFlow library developed by Google's machine intelligence research group \cite{Abadi2016TensorFlow}. This library is open source, and is being extended and improved upon by an active development community. TensorFlow supports GPU and large-scale parallel learning, critical for high-performance optimization. The simple interface to Python allows non-software professionals to implement high-performance machine learning and optimization applications without excessive overhead. 

Typical machine-learning applications require most of the same building blocks needed for quantum optimal control. Predefined operations, along with corresponding gradients, include matrix addition and multiplication; matrix traces; and vector dot products. In addition, we have implemented an efficient kernel for approximate evaluation of the matrix exponential and its gradient. Using these building blocks, we have developed a fast and accurate implementation of quantum optimal control, well-suited to deal with a broad range of engineered quantum systems and realistic treatment of capabilities and limitations of control fields.
 
In common applications of quantum optimal control, time evolving the system under the Schr\"odinger equation -- more specifically, approximating the matrix exponential for the propagators $U_j$ at each time step $t_j$ -- requires the biggest chunk of computational time. Within our matrix-exponentiation kernel, we approximate $e^{-iH_j\delta t}$ by series expansion, taking into account that the order of the expansion plays a crucial role in maintaining accuracy and unitarity. The required order of the matrix-exponential expansion generally depends on the magnitude of the matrix eigenvalues relative to the size of the time step. General-purpose algorithms such as \textsf{expm()} in Python's SciPy framework accept arbitrary matrices $M$ as input, so that the estimation of the spectral radius or matrix norm of $M$, needed for choosing the appropriate order in the expansion, often costs more computational time than the final evaluation of the series approximation itself. Direct series expansion with only a few terms is sufficient for $H_j\delta$ with spectral radius smaller than $1$. In the presence of large eigenvalues, series convergence is slow, and it is more efficient to employ an appropriate form of the ``scaling and squaring'' strategy, based on the identity
\begin{equation}
\exp{M}=\left[\exp\left( \frac{M}{2^n}\right)\right]^{2^n},
\end{equation}
which reduces the spectral range by a factor of $2^n$ at the cost of recursively squaring the matrix $n$ times \cite{Moler2006Nineteen}. Overall, this strategy leads to an approximation of the short-time propagator of the form
\begin{equation}
    U_j\approx \left[\sum_{k=0}^p \frac{(-i H_j \delta t/2^n)^k}{k!}\right]^{2^n},
\end{equation}
based on a Taylor expansion truncated at order $p$. Computational performance could be further improved by employing more sophisticated series expansions \cite{Arioli1996Pade,Cody1969Chebyshev} and integration methods \cite{jameson1981numerical}.

As opposed to the challenges of general-purpose matrix exponentiation, matrices involved in a specific quantum control application with bounded control field strength $(iH_j\delta t)$, will typically exhibit similar spectral radii. Thus, rather than attempting to determine individual truncation levels $p_j$, and performing scaling-and-squaring at level $n_j$ in each time step $t_j$, we make a conservative choice for global $p$ and $n$ at the beginning and employ them throughout. This simple heuristic speeds up matrix exponentiation over the default SciPy implementation significantly, primarily due to leaving out the step of spectral radius estimation.

By default, automatic differentiation would compute the gradient of the approximated matrix exponential via backpropagation through the series expansion. However, for sufficiently small spectral radius of $M$, we may approximate \cite{Khaneja2005Optimal}
\begin{equation}\label{eq:expdiff}
\frac{d}{dx}e^{M(x)} \approx M'(x)\,e^{M(x)},
\end{equation}
neglecting higher-order corrections reflecting  that $M'(x)$ and $M(x)$ may not commute. (Higher-order schemes taking into account such additional corrections are discussed in Ref.\ \onlinecite{deFouquieres2011Second}.)
Equation \eqref{eq:expdiff} simplifies automatic differentiation: within this approximation, only the same matrix exponential is needed for the evaluation of the the gradient. We make use of this in a custom routine for matrix exponentiation and gradient-operator evaluation, further improving the speed and memory performance. 

The TensorFlow library currently has one limitation relevant to our implementation of a quantum optimal control algorithm. 
Operators and states in Hilbert space have natural respresentations as matrices and vectors which are generically complex-valued. TensorFlow, designed primarily for neural network problems, has currently only limited support for complex matrices. For now, we circumvent this obstacle by mapping complex-valued matrices to real matrices via the isomorphism $H \xmapsto{\cong} \openone\otimes H_\mathrm{re} - i\sigma_y\otimes H_\mathrm{im}$, and state vectors $\vec{\Psi} \xmapsto{\cong} (\vec{\Psi}_\mathrm{re}, \vec{\Psi}_\mathrm{im})^t$. Here, $\openone$ is the 2$\times$2 unit matrix and $\sigma_y$ one of the Pauli matrices. Real and imaginary part of the matrix $H$ are denoted by $H_\mathrm{re}=\rre H$ and $H_\mathrm{im}=\iim H$, respectively; similarly, real and imaginary parts of state vectors are $\vec{\Psi}_\mathrm{re}=\rre \vec{\Psi}$ and $\vec{\Psi}_\mathrm{im}=\iim \vec{\Psi}$. Written out in explicit block matrix form, this isomorphism results in 
\begin{equation}
H \vec{\Psi} \xmapsto{\cong}
\begin{pmatrix}
    H_\mathrm{re}      & -H_\mathrm{im} \\
    H_\mathrm{im}       & H_\mathrm{re} 
\end{pmatrix}
\begin{pmatrix}
    \vec{\Psi}_\mathrm{re}     \\
    \vec{\Psi}_\mathrm{im}  
\end{pmatrix},
\label{eq:matrix_isomorphism}
\end{equation}
rendering all matrices and vectors real-valued. For the Hamiltonian matrix, this currently implies a factor two in memory cost (due to redundancy of real and imaginary part entries). There are promising indications that future TensorFlow releases may improve complex-number support and eliminate the need for a mapping to real-valued matrices and vectors.

\section{Performance Benchmarking \label{sec:performance_benchmarking}}

\begin{figure*}
  \centering \includegraphics[width=0.9\textwidth]{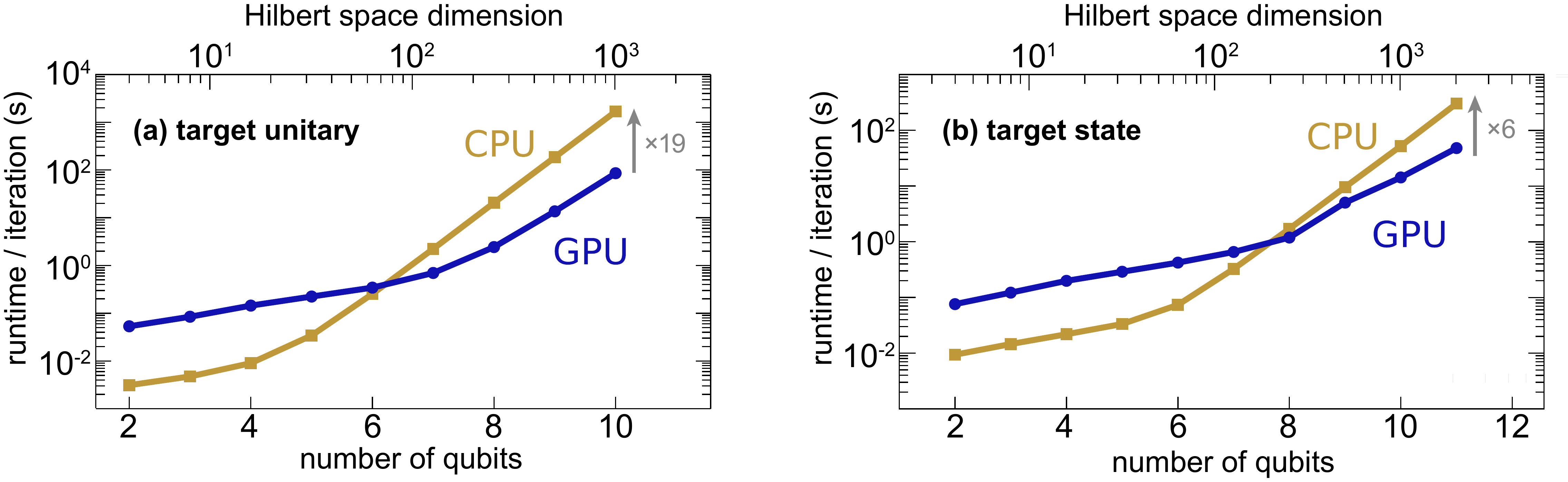}
  \caption{Benchmarking comparison between GPU and CPU for (a) a unitary gate (Hadamard transform), and (b) state transfer (GHZ state preparation). Total runtime per iteration scales linearly with the number of time steps. For unitary-gate optimization, the GPU outperforms the CPU for Hilbert space dimensions above $\sim100$. For state transer, GPU benefits set in slightly later, outperforming the CPU-based implementation for Hilbert space dimensions above $\sim 300$. The physical system we consider, in this case, is an open chain of $N$ spin-1/2 systems with nearest neighbor $\sigma_z \sigma_z$ coupling, and each qubit is controlled via fields $\Omega_x$ and $\Omega_y$.}
  \label{fig:benchmark}
\end{figure*}

Obtaining a fair comparison between CPU-based and GPU-based computational performance is notoriously difficult \cite{Lee2010Debunking}. We attempt to provide a specific comparison under a unified computation framework. TensorFlow allows for straightforward switching from running code on a CPU to a GPU. For each operation (matrix multiplication, trace, etc.), we use the default CPU/GPU kernel offered by TensorFlow. Note that properly configured, TensorFlow automatically utilizes all threads available for a given CPU, and GPU utilization is found to be near 100\%. Not surprisingly, we observe that the intrinsic parallelism of GPU-based matrix operations allows much more efficient computation beyond a certain Hilbert space size, see Fig.\ \ref{fig:benchmark}. 

% The physical system, in this case, is an open chain of $N$ spin-1/2 systems, described by

% \begin{equation}
%     H(t) = \sum_{n=1}^N \left[ \Omega^{(n)}_x (t) \sigma_x^{(n)} + \Omega_y^{(n)} (t) \sigma_y^{(n)}\right] + \sum_{n=1}^{N-1} J\,\sigma_z^{(n)} \sigma_z^{(n+1)},
% \end{equation}
% \begin{equation}
% \mathcal{H}_0 = \sum_{\ell=1}^N \frac{\epsilon}{2}\sigma_\ell^z + \sum_{\langle \ell, \ell'\rangle} \sigma_\ell^x\sigma_{\ell'}^x.
% \end{equation}
% Beyond this system Hamiltonian $\mathcal{H}_0$, we assign a single, global control field to the joint operator $\sigma_x\sigma_x\dots\sigma_x$. This choice, though somewhat artificial, is motivated by separating the influence of the number of control fields from the number of spins, and hence Hilbert space dimension. The full Hamiltonian then reads $H(t) = \mathcal{H}_0 + u(t) \sigma_x^{(1)}\sigma_x^{(2)}\cdots\sigma_x^{(N)}$. While not related to a particular physical realization, this Hamiltonian is very suitable for benchmarking purposes. \textcolor{orange}{[Though the NL update was important, it does not address my initial request: NEED INFORMATION ON WHAT GATE/WHAT STATE YOU ARE OPTIMIZING FOR]} 

In this example, we specifically inspect how the computational speed scales with the Hilbert space dimension when optimizing an $n$-spin Hadamard transform gate and $n$-spin GHZ state preparation for a coupled chain of spin-1/2 systems presented in Section \ref{sec:spin_chain_example}. (Details of system parameters are described in the same section.) We benchmark the average runtime for a single iteration for various spin-chain sizes and, hence, Hilbert space dimensions. We find that the GPU quickly outperforms the CPU in the unitary gate problem, even for a moderate system size of $\sim 100$ basis states. For optimization of state transfer, we observe that speedup from GPU usage, relative CPU performance, sets in for slightly larger system sizes of approximately $\sim 300$ basis states. 

The distinct thresholds for the GPU/CPU performance gain stem from the different computational complexities of gate vs.\ state-transfer optimization. Namely, 
optimizing unitary gates requires the propagation of a unitary operator (a matrix), involving matrix-matrix multiplications, while optimizing state transfer only requires the propagation of a state (a vector), involving only matrix-vector multiplications: 
\begin{equation}
    U_j|\Psi\rangle \approx \sum_{k=0}^p \frac{(-i \delta t)^k}{k!} (H_j\dots(H_j(H_j |\Psi\rangle))),
\end{equation}
Computing the matrix-vector multiplication is generally much faster than computing the matrix exponential itself \cite{Sidje1998Expokit}.  For an $n$-dimensional matrix, the computation of the matrix exponential involves matrix-matrix multiplication, which scales as $O(n^3)$. The computation of state transfer only involves matrix-vector multiplication, which scales as $O(n^2)$ [or even $O(n)$ for sufficiently sparse matrices]. 

For optimization of the Hadamard transform as well as the GHZ state preparation, we observe a 19-fold GPU speedup for a 10-qubit system (Hilbert space dimension of 1,024) in the former case, and a 6-fold GPU speedup for an 11-qubit system (Hilbert space dimension of 2,048) in the latter case. Since matrix operations are the most computationally intensive task in our software, this speedup is comparable to other GPU application studies that heavily use matrix operation \cite{Oh2004GPU,Catanzaro2008Fast,Sharp2008Implementing,Raina2009Largescale,Steinkraus2005Using,2016arXiv160502688short,Lee2010Debunking}.
We emphasize that these numbers are indicative of overall performance trends, but detailed numbers will certainly differ according to the specific system architecture in place. The CPU model we used was an Intel\textsuperscript{\textregistered} Core\textsuperscript{\texttrademark } i7-6700K CPU @ 4.00\,GHz, and the GPU model was an NVIDIA\textsuperscript{\textregistered} Tesla\textsuperscript{\textregistered} K40c. In this study, all computations are based on dense matrices. Since most physically relevant Hamiltonians are sparse (evolution generally affects sparsity, though), future incorporation of sparse matrices may further improve  computation speed for both CPU and GPU \cite{Bell:SpMV:NVIDIA:2008,Liu2014Efficient}. 

\begin{figure*}
  \centering \includegraphics[width=0.9\textwidth]{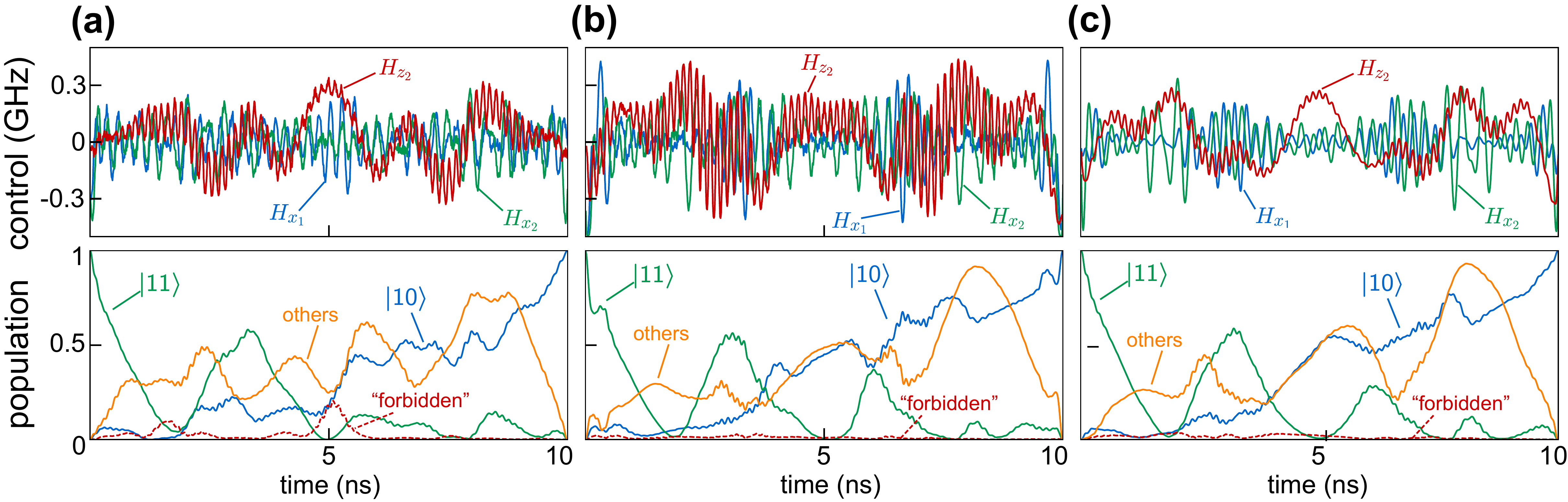}
  \caption{Control pulses and evolution of quantum state population for a CNOT gate acting on two transmon qubits, (a) only targeting the desired final unitary, (b) employing an additional cost function suppressing occupation of higher-lying states ($C_5$), and (c) including additional pulse-shape cost functions ($C_3, C_4$). Here, only the evolution of state $|11\rangle$ is shown, as the evolution of state $|11\rangle$ is most susceptible to the occupation of higher level states. In all three cases, the CNOT gate converged to a fidelity of 99.9\%. The results differ in important details: in (a), both high-frequency ``noise'' on the control signals and significant occupation of ``forbidden'' states (3rd and 4th excited transmon level), shown as dashed red line, are visibile throughout the evolution; in (b), forbidden-state occupation is suppressed at each time step during evolution; in (c), this suppression is maintained and all control signals are smoothened. The maximum occupation of forbidden states is reduced from $\sim 20\%$ in (a) to $\sim 3\%$ in (b) and (c). The population of ``others'' states (non- $|11\rangle$, $|10\rangle$ or ''forbidden'') is also shown for completeness. For demonstration purposes, all three examples use the same gate duration of $10\,\text{ns}$, despite being subject to different constraints. In practice, one would typically increase the gate time for a more constrained problem to achieve the best result in maximizing gate fidelity, minimizing forbidden state occupation, and achieving a realistic control signal. }
  \label{figure:transmon_transmon_cnot}
\end{figure*}

\section{Showcase Applications  \label{sec:examples}}
In this last section, we present a set of example applications of experimental relevance. The first application demonstrates the importance of cost functions suppressing intermediate occupation of higher-lying states during time evolution, as well as cost functions accounting for realistic pulse shaping capabilities. In a second application, we show how the cost function $C_6$ can yield high-fidelity state transfer within a reduced time interval. Third, we discuss the application of Schr\"odinger-cat state preparation -- an example from the context of quantum optics and of significant interest in recent schemes aiming at quantum information processing based on such states \cite{Heeres2016Implementing,Mirrahimi2013Dynamically,Vlastakis2013Deterministically}. This application combines considerable system size with a large number of time steps, and utilizes most of the cost functions discussed in Section \ref{sec:costfunctions}. In the fourth application, we demonstrate the algorithm performance in finding optimal solutions for GHZ state preparation and implementation of a Hadamard transform gate in a chain of qubits with a variable number of qubits. We use either the Adam \cite{Kingma2015Adam} or L-BFGS-B optimization algorithm \cite{Byrd1995Limited} for pulse optimization, and achieve a minimum fidelity of 99.9$\%$ in all of our following examples.

\subsection{CNOT gate for two transmon qubits}
In the first example, we study realization of a 2-qubit CNOT gate in a system of two coupled, weakly anharmonic transmon qubits. For each transmon qubit $(j=1,2)$ \cite{Koch2007Chargeinsensitive}, we take into account the lowest two states spanning the qubit computational space, as well as the next three higher levels. 
The system Hamiltonian, including the control fields $\{\Omega_{x_1}(t),\,\Omega_{x_2}(t),\,\Omega_{z_2}(t)\}$, then reads
\begin{align}
    H(t) &= \sum_{j=1,2} \left[\omega_j b_j^\dag b_j + \tfrac{1}{2} \alpha_j\,b_j^\dag b_j (b_j^\dag b_j-1)\right]\\\nonumber
    &\quad 
    + J(b_{1}+b_{1}^{\dagger})(b_{2}+b_{2}^{\dagger})\\\nonumber
    &\quad +  \Omega_{x_1} (t)(b_{1}+b_{1}^{\dagger})     +  \Omega_{x_2} (t)(b_{2}+b_{2}^{\dagger}) + \Omega_{z_2} (t)b_{2}^{\dagger}b_{2}.
\end{align}
Here, the ladder operators $b_j$, and $b_j^\dag$ are truncated at the appropriate level. (The qubit frequencies $\omega_j/2\pi$ are chosen as 3.5 and 3.9\,GHz, respectively; both transmons have an anharmonicity of $\alpha/2\pi=-225\,\text{MHz}$; and the qubit-qubit coupling strength used in the simulation is $J/2\pi=100\,\text{MHz}$.)
Consistent with recent circuit QED experiments utilizing classical drives as well as parametric modulation, we investigate control fields acting on $H_{x_1}=b_{1}+b_{1}^{\dagger},  \, H_{x_2}=b_{2}+b_{2}^{\dagger}$, and  $H_{z_2}=b_{2}^{\dagger}b_{2}$. 

We next optimize control fields for the realization of a CNOT gate, with transmon qubit $j=1$ acting as the control qubit. Our control-field optimization reaches a prescribed fidelity of 99.9$\%$ for a $10\,\text{ns}$ gate duration in all cases, as seen in Fig.\ \ref{figure:transmon_transmon_cnot}. Results shown in Fig.~\ref{figure:transmon_transmon_cnot}(a) are obtained with the standard target-gate infidelity cost function ($C_1$) only. It is evident that the solution encounters two issues: the occupation of the 3rd and 4th excited transmon level (``forbidden'') is significant, and control fields are polluted by high-frequency components. Including a cost function contribution of type $C_5$ succeeds in strongly suppressing occupation of higher levels, see Fig.~\ref{figure:transmon_transmon_cnot}(b). This both reduces exposure to increased relaxation rates and ensures that the evolution is minimally influenced by our numerical truncation of Hilbert space. In the final improvement step, shown in Fig.~\ref{figure:transmon_transmon_cnot}(c), our optimization additionally suppresses excessive control amplitudes and derivatives via cost contributions of type $C_3$ and $C_4$. The inclusion of these terms in the overall cost lessens superfluous ``noise'' in the control signals, and also helps improve convergence of the algorithm -- without reducing the achieved target-gate fidelity.

\begin{figure}[ht]
  \centering \includegraphics[width=0.9\columnwidth]{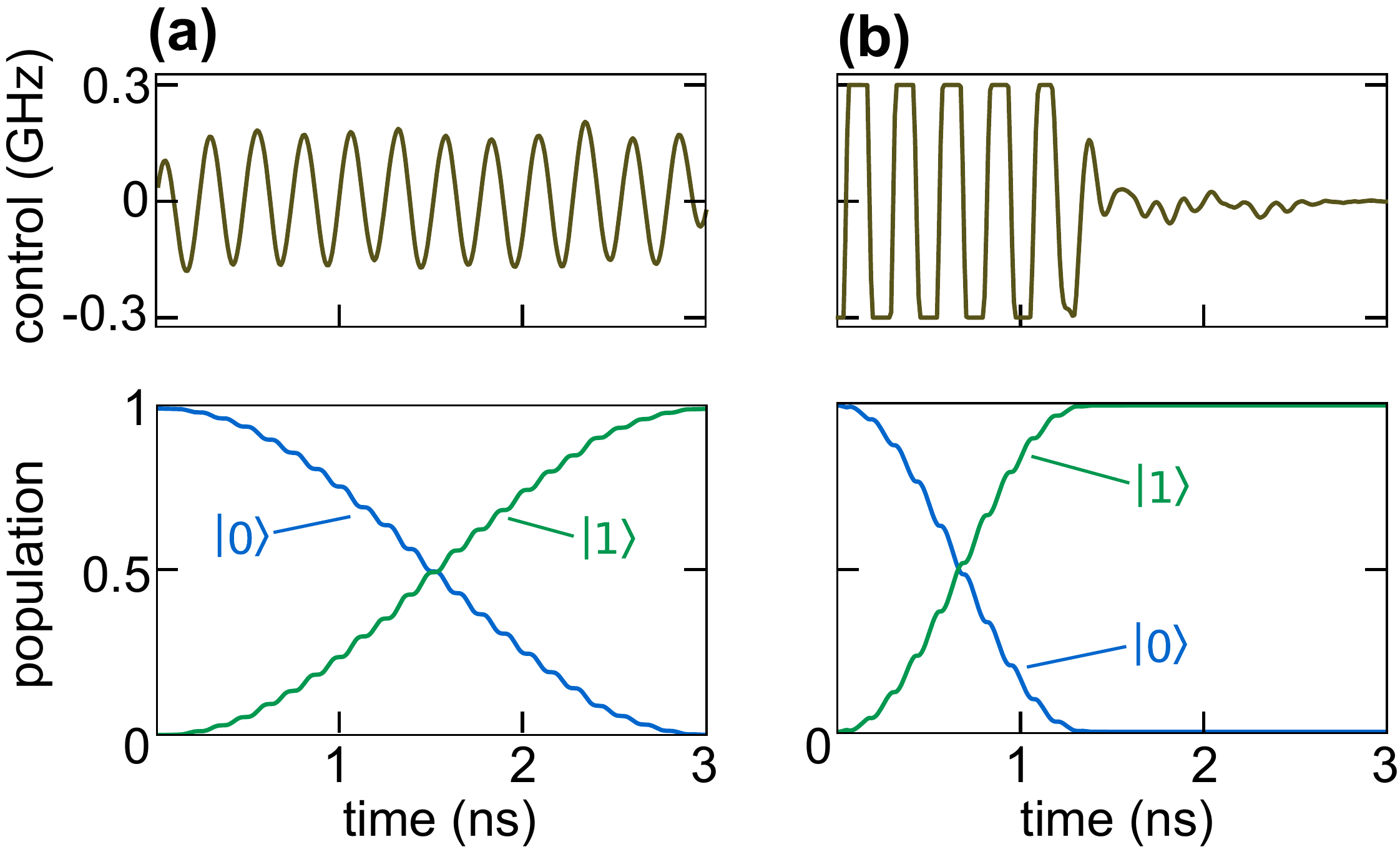}
  \caption{{Minimizing evolution time needed for a high-fidelity state transfer. (a) No time-optimal award function. (b) With time-optimal award function.}
  (a) Without penalty for the time required for the gate, the control field spreads across the entire given time interval. (b) Once evolution over a longer time duration is penalized with a contribution of type $C_6$ or $C_7$ (see table I), the optimizer achieves target state preparation in a shorter time, without loss of fidelity.}
  \label{figure:spin_pi_speed_up}
\end{figure}

\subsection{Reducing duration of $|0\rangle$ to $|1\rangle$ state transfer}
In this second example, we illustrate the use of cost function contributions (types $C_6$, $C_7$) in minimizing the time needed to perform a specific gate or prepare a desired state. To this end, we consider a two-level spin qubit ($\omega/2\pi$: $3.9\,\text{GHz}$). The system and control Hamiltonians combined are taken to be 
\begin{equation}
    H = \frac{\omega}{2} \sigma_z + \Omega (t)\,\sigma_x.
\end{equation}
We allow for a control field acting on the qubit $\sigma_x$ degree of freedom, and constrain the maximum control-field strength $\Omega_\text{max}/2\pi$ to $300\,\text{MHz}$. When the evolution time needed to perform the state transfer is fixed (rather than subject to optimization itself), we observe that control fields generically spread across the prescribed gate duration time. The desired target state is realized only at the very end of the allowed gate duration. When we incorporate a $C_6$ or $C_7$-type cost contribution, the optimal control algorithm also aims to minimize the overall gate duration, so as to realize the target unitary or state in as short a time as possible, given other active constraints. In our example, this reduces the time for a state transfer from $3\,\text{ns}$ to less than $1.5\,\text{ns}$, see Fig.\ \ref{figure:spin_pi_speed_up}. We note that it is further possible to adaptively change the overall simulation time during optimization. For instance, if further optimization was desired in the case of Fig.\ \ref{figure:spin_pi_speed_up}(b), then the simulation time interval could be adaptively reduced to $\sim1.5\,\text{ns}$ -- resulting in a significant cutback in overall computation time.

\begin{figure*}
  \centering \includegraphics[width=1.0\textwidth]{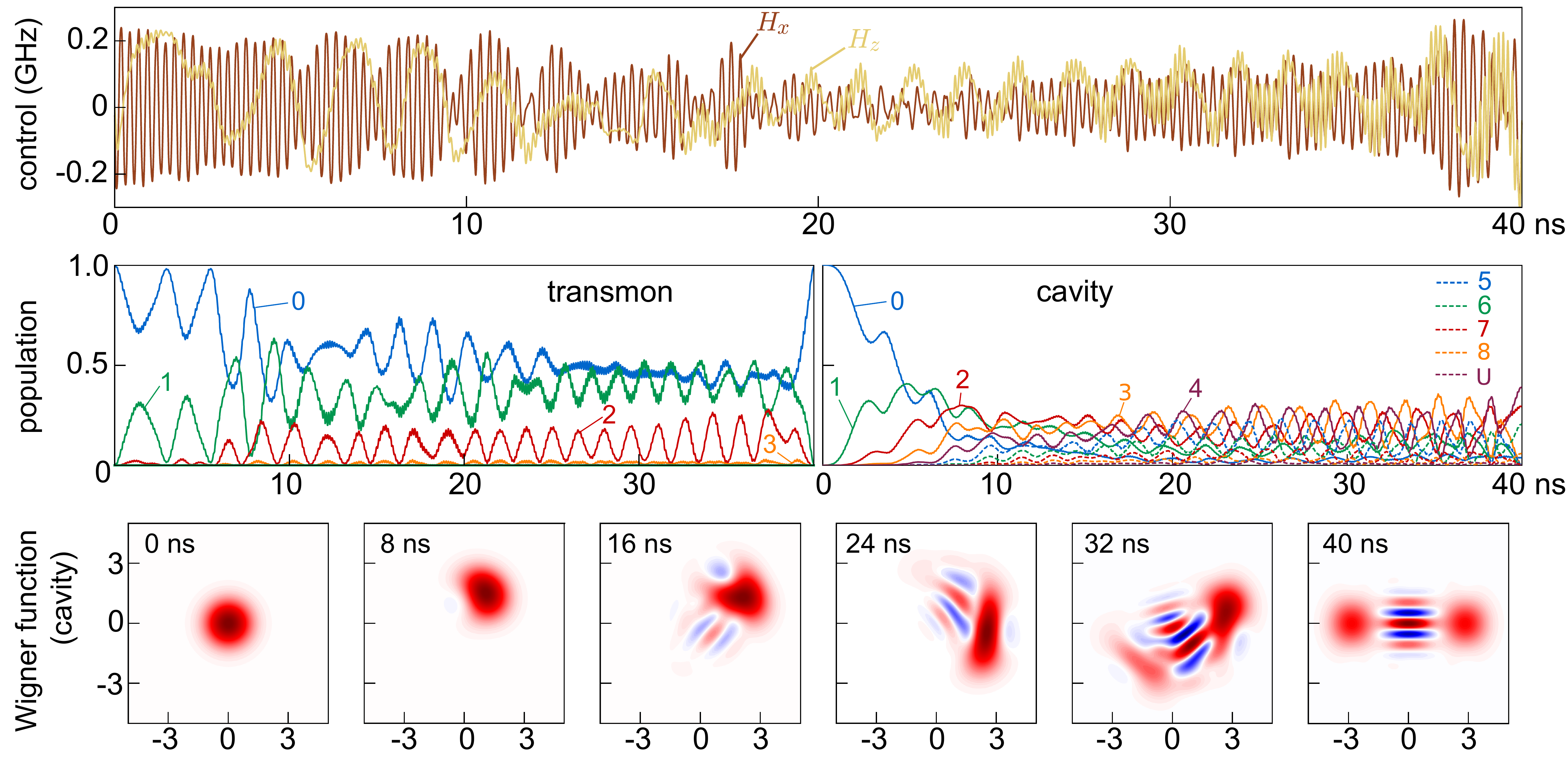}
  \caption{{Cat state generation. Control pulse, state evolution in Fock basis, and Wigner function tomography of the cavity evolution.} Photonic cat state generation is shown as a test of state transfer, challenging the quantum control algorithm with a system of considerable size, large number of required time steps, and inclusion of multiple types of cost function. The desired Schr\"odinger cat state in the resonator is created indirectly, by applying control fields to a transmon qubit coupled to the resonator, and reached within a prescribed evolution time of $40\,\text{ns}$ with a fidelity of 99.9\%. (Note that occupation of transmon level 4, 5, 6 remains too small to be visible in the graph.) 
  \label{figure:cat_state}}
\end{figure*}
\subsection{Generating photonic Schr\"odinger cat states}
As an example of quantum state transfer, we employ our optimal control algorithm to the task of generating a photonic Schr\"odinger-cat state. The system we consider to this end is a realistic, and recently studied \cite{Mirrahimi2013Dynamically,Vlastakis2013Deterministically} circuit QED setup, consisting of a transmon qubit capacitively coupled to a three-dimensional microwave cavity. External control fields are restricted to the qubit. Working in a truncated subspace for the transmon (limiting ourselves to levels with energies well below the maximum of the cosine potential), the full Hamiltonian describing the system is
\begin{align}
    H(t) &= \omega_q b^\dag b +\tfrac{1}{2} \alpha\,b^\dag b (b^\dag b-1)  + \omega_r a^\dag a\\\nonumber
    &\quad + g(a+a^{\dagger})(b+b^{\dagger}) +  \Omega_{x} (t)(b+b^{\dagger}) 
     +  \Omega_{z} (t)b^{\dagger}b
\end{align}
Here, $a$ and $b$ are the usual lowering operators for photon number and transmon excitation number, respectively. The frequencies $\omega_q/2\pi = 3.5\,\text{GHz}$ and $\alpha/2\pi=-225\,\text{MHz}$ denote the transmon 0-1 splitting and its anharmonicity. The frequency of the relevant cavity mode is taken to be $\omega_r/2\pi=3.9\,\text{GHz}$. Qubit and cavity are coupled, with a strength parameterized by $g/2\pi=100\,\text{MHz}$. In our simulation, the overall dimension is $154=(\text{7 transmon levels})\times (\text{22 resonator levels})$. Note that the rotating wave approximation is not applied in order to reflect the capabilities of modern arbitrary waveform generation.

The state-transfer task at hand, now, is to drive the joint system from the zero-excitation state $|0\rangle_q\otimes|0\rangle_r$ (the ground state if counter-rotating terms in the coupling are neglected) to the photonic cat state $|0\rangle_q\otimes|\text{cat}\rangle_r$. Here, the cat state in the resonator corresponds to a superposition of two diametrically displaced coherent states: $|\text{cat}\rangle_r = \frac{1}{\sqrt{2}} (|\lambda\rangle + |-\lambda\rangle)$. Coherent states are defined in the usual way as normalized eigenstates of the photon annihilation operator $a$, and correspond to displaced vacuum states $|\lambda\rangle= e^{-|\lambda|^2/2} e^{\lambda a^\dagger}|0\rangle$. The cat state $|\text{cat}\rangle_r$ is approximately normalized for sufficiently large $\lambda$. As our concrete target state, we choose a cat state with amplitude $\lambda=2$ (normalization error of $\sim$ 0.03\%). The state transfer is to be implemented by control fields $\Omega_x(t)$ and $\Omega_z(t)$ acting on the transverse and longitudinal qubit degrees of freedom, $H_{x}=(b+b^{\dagger})$ and $H_{z} = b^{\dagger}b$, respectively. Matching experimental realizations of multi-mode cavity QED systems\cite{McKay2015HighContrast}, we do not allow for any direct control of the cavity degrees of freedom.

This state-transfer problem provides an excellent test for an optimal control algorithm. It incorporates the simultaneous challenges of a large number of time steps (8,000), a considerable evolution time ($40\,\text{ns}$), and the application of most of the cost functions we discussed in Sect.\ \ref{sec:costfunctions} and summarized in Table \ref{table:cost_functions}. Specifically, in addition to minimizing the target state infidelity ($C_2$), we penalize occupation of transmon levels 3 to 6 and cavity levels 20 and 21 ($C_5$) to avoid artifacts from truncation, and penalize control variations ($C_4$) \footnote{A cost function for reducing evolution time ($C_7$) was not included in this example.}. Results from the optimization are presented in \ref{figure:cat_state}, which shows the control-field sequence, as well as the induced state evolution. At the end of the $40\,\text{ns}$ time interval, the control fields generate the desired cat state with a fidelity of 99.9\%. The maximum populations at the truncation levels of transmon and cavity are $\sim 6\times 10^{-6}$ and $\sim 7\times 10^{-10}$, respectively. We independently confirm convergence with respect to truncation by simulating the obtained optimized pulse for enlarged Hilbert space  (8 transmon and 23 cavity levels), and find that the evolution continues to reach the target state with 99.9\% fidelity.

\subsection{Hadamard transform and GHZ state preparation\label{sec:spin_chain_example}}
We present a final set of examples illustrating the algorithm performance for increasing system size. To that end, we consider a coupled chain of $N$ qubits, or spin-1/2 systems. We assume that all spins are on-resonance in the multiple-rotating frame. This system is described by the Hamiltonian
\begin{equation}
    H(t) = \sum_{n=1}^N \left[ \Omega^{(n)}_x (t) \sigma_x^{(n)} + \Omega_y^{(n)} (t) \sigma_y^{(n)} + J\,\sigma_z^{(n)} \sigma_z^{(n+1)} \right],
\end{equation}
where the coupling term is understood to be dropped for the final summand ($n=N$). 
The qubit-qubit coupling strength is fixed to $J/2\pi =100$\,MHz. Each qubit $(n)$ is controlled via fields $\Omega_x^{(n)}$ and $\Omega_y^{(n)}$, with a maximum allowed drive strength of $\Omega^{(n)}_{x,y}/2\pi = 500$\,MHz.

\begin{figure}
  \centering \includegraphics[width=0.95\columnwidth]{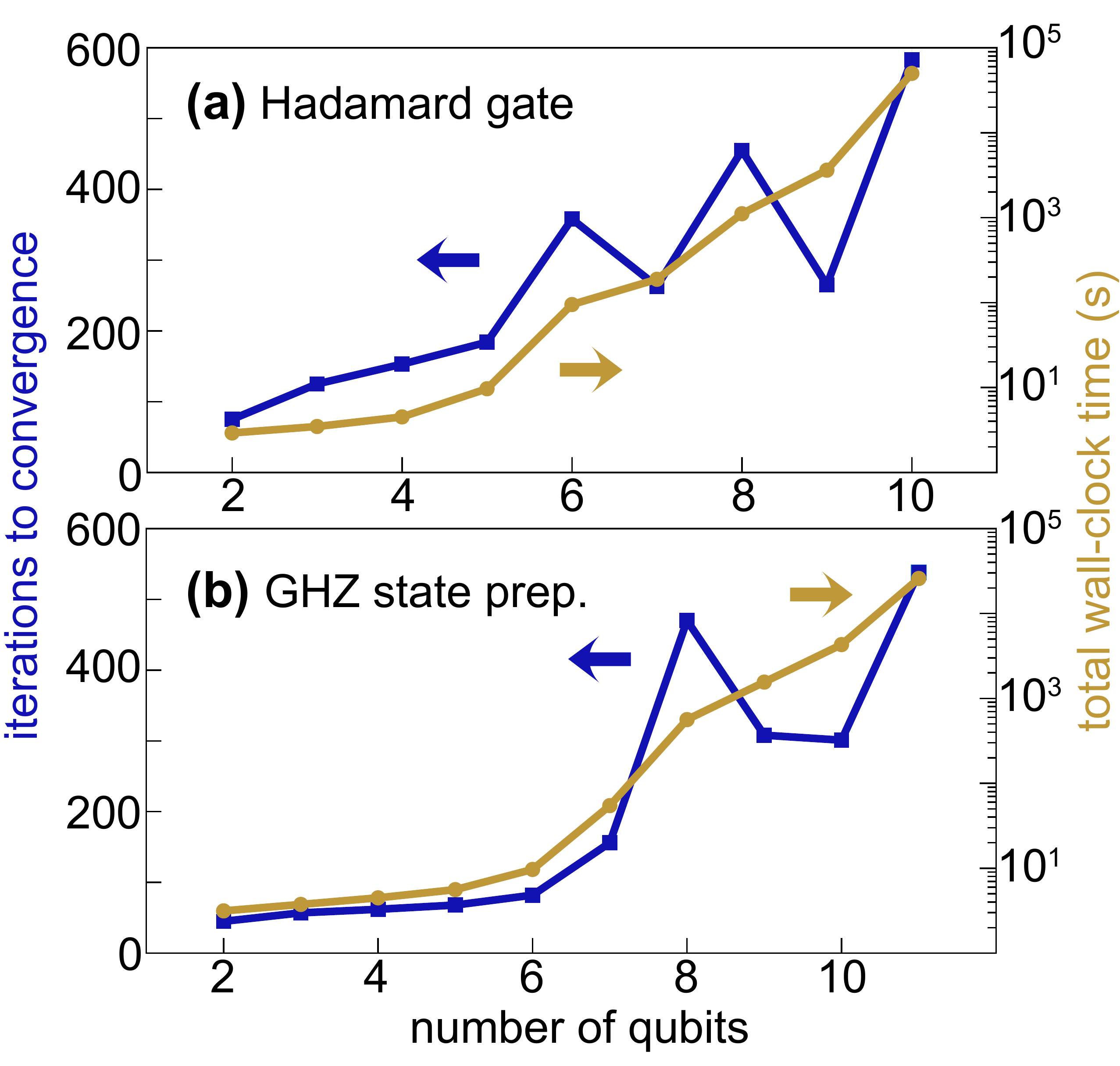}
  \caption{Performance of optimal control algorithm as a function of qubit number for (a) a Hadamard transform gate, and (b) GHZ state preparation. As system size increases, total time and number of iterations for the algorithm grow rapidly. The larger number of control parameters and complexity of the target state add to the challenge of quantum optimal control for systems with many degrees of freedom.}
  \label{figure:spin_chain_ghz_hadamard}
\end{figure}

As a first optimization task, we search for control fields to implement the unitary realizing a Hadamard transform, commonly used in various quantum algorithms. The gate time we allow for the Hadamard transform is $(2N)\,\text{ns}$, simulated with $10N$ time steps. Figure \ref{figure:spin_chain_ghz_hadamard}(a) shows the number of iterations and wall-clock time required to converge to the desired 99.9\% process fidelity. For the same spin-chain system, we have also employed our code to optimize control fields for transferring the system ground state to a maximally entangled GHZ state. The overall time and time steps we allow for the GHZ state preparation is identical to that used  for the Hadamard transform gate. Figure \ref{figure:spin_chain_ghz_hadamard}(b) shows the number of iterations necessary and the total wall-clock time spent for reaching convergence to a result with 99.9\% state fidelity. For both examples, we employed computation on either CPU or GPU, depending which one is faster. (This performance benchmarking data was shown in Section \ref{sec:performance_benchmarking}). We note that, when using a modest desktop PC with graphics card, optimal control problems for small Hilbert space size converge within seconds. For a 10-qubit Hadamard gate (Hilbert space dimension of 1,024) or 11-qubit GHZ state (Hilbert space dimension of 2048), it takes $\sim$1 day to obtain a solution meeting the 99.9\% fidelity threshold. The total wall-clock time could likely have been reduced significantly by appropriate choice of optimizer, hyperparameters, and/or initial control fields. In the case of spin-chain system, like many quantum information systems, as the number of elements increase, not only does the Hilbert space grow exponentially, the number of control fields and the required number of time steps also get larger. This further increases the complexity of the problem.

\section{Conclusion}
In conclusion, we have presented a quantum optimal control algorithm harnessing two key technologies that enable fast and low-overhead numerical exploration of control signal optimization. First, we have demonstrated that automatic differentiation can be leveraged to facilitate effortless inclusion of diverse optimization constraints, needed to obtain realistic control signals tailored for the specific experimental capabilities at hand. Automatic differentiation dramatically lowers the overhead for adding new cost functions, as it renders analytical derivations of gradients unnecessary. 
For illustration, we have presented concrete examples of optimized unitary gates and state transfer, using cost functions relevant for applications in superconducting circuits and circuit QED. We emphasize that this is but one instance within a much larger class of quantum systems for which optimal control is instrumental, and the methods described here are not limited to the specific examples shown in this paper.

The second key technology we have incorporated is the implementation of GPU-based numerical computations, which offers a significant speedup relative to conventional CPU-based code. The use of the TensorFlow library \cite{Abadi2016TensorFlow} hides the low-level details of GPU acceleration, allowing implementation of new cost functions at a high level. The reduction in computational time will generally depend on a number of factors including system type, Hilbert space size, and the specific hardware employed by the user. We observe that runtime speedup by an order of magnitude is not unusual when using a standard desktop PC, enabling the development of sophisticated quantum control without enormous investments into powerful computing equipment.  The underlying libraries also have support for high-performance distributed computing systems for larger optimizations. Our software implementation is open source and can be downloaded at: \url{github.com/SchusterLab/quantum-optimal-control}. 

The increased efficiency and ease of optimal quantum control due to the employment of GPUs and automatic differentiation makes our work valuable to a broad range of research. Future work will address sparse-matrix implementations, as well as the deployment of adaptive step size and Runge-Kutta methods for time evolution.

\begin{acknowledgments}
The authors thank J.\ Kutasov for help in comparing optimization performance with existing frameworks as well as T.\ Berkelbach and D.\ Mazziotti for discussions.
We gratefully acknowledge the support of NVIDIA\textsuperscript{\textregistered} Corporation through the donation of the Tesla\textsuperscript{\textregistered} K40 GPU used for this research, and support from the David and Lucille Packard Foundation. This material is based upon work supported by the Department of Defense under contracts H98230-15-C0453 and W911NF-15-1-0421.
\end{acknowledgments}

%\clearpage
\appendix
\section{Analytical gradients and algorithms \label{app:gradient}}
In the following, we outline the analytical calculation of gradients for cost functions such as those summarized in Table \ref{table:cost_functions}. We stress that our automatic-differentiation implementation evaluates these gradients autonomously, without the need of these analytical derivations or hard-coding any new gradients. The following derivations are thus merely intended as illustrations for a better mathematical understanding (and appreciation) of the gradients calculated without user input by means of automatic differentiation.

For a systematic treatment of the different types of cost functions, we note that most cost functions involve an absolute value squared of an inner product between target and final states or target and final unitaries (Hilbert-Schmidt inner product). To obtain the gradients of expressions such as $C_1(\mathbf{u}) =  1- |\tr(K_T^\dagger K_{N})|^2$ with respect to the control parameters, we note that control parameters enter via the final states or unitaries through the evolution operators, $K_{N}=U_N(\mathbf{u})U_{N-1}(\mathbf{u})\cdots U_1(\mathbf{u})U_0$. 
To streamline our exposition, we first summarize multiple matrix-calculus relations of relevance. 

Consider two complex-valued matrices $A$ and $B$, compatible in row/column format such that the matrix product $AB$ is defined. Then, one finds
\begin{equation}
\frac{\partial \tr(AB)}{\partial B_{ji}} =  \frac{\partial (A_{nm}B_{mn})}{\partial B_{ji}} = A_{ij}.
\label{eq:trace_calculus_1}
\end{equation}
Throughout this appendix, we use Einstein convention for summation, and follow the Jacobian formulation (also known as numerator layout) for derivatives with respect to matrices.
%
%Therefore,
%\begin{equation}
%\begin{align}
%\begin{split}
%\frac{\partial \tr(AB)}{\partial B}  &= A\\ 
%\tr(\frac{\partial \tr(AB)}{\partial B} \delta B)  &= \tr(A \delta B)
%\end{split}
%\end{align}
%\label{eq:trace_calculus_2}
%\end{equation}
%
We will further encounter expressions of the following form, involving a third matrix $C$ of the same dimensions as $B^t$: 
\begin{align}\nonumber
&\tr\bigg[\frac{\partial [|\tr(AB)|^2]}{\partial B} C\bigg]%\\\nonumber
%&\qquad=\frac{\partial (|\tr(AB)|^2)}{\partial B_{ji}} C_{ji}%\\\nonumber
%&\quad
= \frac{\partial [\tr(AB) \tr(AB)^*]}{\partial B_{ji}} C_{ji}\\\nonumber
&\qquad= \frac{\partial \tr(AB)}{\partial B_{ji}} \tr(AB)^* C_{ji}
= A_{ij}  \tr(AB)^* C_{ji}\\
&\qquad = \tr (A C) \tr(AB)^*.
%\frac{\partial \{ [Re(tr(AB))]^2 + [Im(tr(AB))]^2 \}}{\partial B} \delta B 
%\\ = 2 Re(tr(AB)) \frac{\partial [Re(tr(AB))]}{\partial B} \delta B +\\ 2 Im(tr(AB)) \frac{\partial [Im(tr(AB))]}{\partial B} \delta B 
%\\ = 2 Re(tr(AB)) Re[\frac{\partial (tr(AB))}{\partial B} \delta B] +\\ 2 Im(tr(AB)) Im[\frac{\partial (tr(AB))}{\partial B} \delta B]
%\\ = 2 Re[tr(AB)] Re[tr(A \delta B)] +\\ 2 Im[tr(AB)] Im[tr(A \delta B)]
%\\ = 2 Re[ tr(A \delta B) tr(AB)^*]
\label{eq:trace_calculus_3}
\end{align}
In the framework of Wirtinger derivatives in complex analysis, derivatives treat quantities $X$ and $X^*$ as independent variables, and Eq.\ \eqref{eq:trace_calculus_1} is used in the step from line 1 to line 2. 

The evaluation of cost-function gradients requires the application of the chain rule to expressions of the type $\frac{\partial}{\partial u_i} c(M(\mathbf{u}))$. Here, $c$ maps a complex-valued $\ell\times \ell$ matrix $M$ (e.g., the propagator $K_N$ with $\ell$ denoting the Hilbert space dimension) to a real number (the cost). The matrix $M=(M_{mn})$ itself depends on the real-valued control parameters $\mathbf{u}\in \mathbb{R}^d$. The subscript in $u_i$ is understood as a multi-index $i=(k,j)$ encoding the control-field label $k$ and discretized-time index $j$. The matrix-calculus result
\begin{align}\nonumber
 \frac{\partial}{\partial u_i} c(M(\mathbf{u})) &= \frac{\partial c}{\partial M_{mn}} \frac{\partial M_{mn}}{\partial u_i} + \frac{\partial c}{\partial M_{mn}^*} \frac{\partial M_{mn}^*}{\partial u_i}\\
&= 
\tr\left(\frac{\partial c}{\partial M} \frac{\partial M}{\partial u_i}\right)
+\text{c.c.}
\label{eq:chain_rule}
\end{align}
is straightforward to derive with the ``regular'' chain rule by re-interpreting the functions involved as $c\colon \mathbb{C}^{\ell^2} \rightarrow \mathbb{R}$ and $M\colon \mathbb{R}^d\rightarrow\mathbb{C}^{\ell^2}$. 
In the following, Eqs.\ \eqref{eq:trace_calculus_3} and \eqref{eq:chain_rule} are used to obtain the analytical expressions for several examples of cost-function gradients.

\subsection{Gradient for $C_1$: target-gate infidelity}
The cost function $C_1 =  1- |\tr[K_T^\dagger K_{N}(\mathbf{u})]/D|^2$, penalizes the infidelity of the realized unitary $K_N=U_N U_{N-1}\ldots U_1 U_0$ with respect to the target propagator $K_T$. In the following, we omit the constant factor $D$ since it affects all the gradients only by a constant factor. The cost function then has the gradient
\begin{align}
\label{eq:target_unitary_gradient_expression}
& \frac{\partial C_1}{\partial u_{k,j}} \overset{\eqref{eq:chain_rule}}= 
\tr \frac{\partial C_1}{\partial K_N}\frac{\partial K_N}{\partial u_{k,j}}+\text{c.c.}\\\nonumber
&\quad=
-\tr \bigg[\frac{\partial[|\tr(K_T^\dagger K_{N})|^2]}{\partial K_N}\frac{\partial K_N}{\partial u_{k,j}}\bigg] +\text{c.c.}
\\\nonumber
&\quad\overset{\eqref{eq:trace_calculus_3}}=
-\tr \bigg( K_T^\dag \frac{\partial K_N}{\partial u_{k,j}}\bigg)
\tr ( K_T^\dag K_N)^* +\text{c.c.}
\\\nonumber
&\quad
=  \tr \bigg(K_T^\dagger
\Big[\prod_{j'>j} U_{j'}\Big] i\, \delta t\, \mathcal{H}_k K_{j}\bigg)\tr(K_T^\dagger K_{N})^* + \text{c.c.}\\\nonumber
&\quad = 
-2\,\delta t \iim\bigg \{
 \tr \bigg(  K_{T}^\dagger \Big[  \prod_{j'>j} U_{j'} \Big]\, \mathcal{H}_k  K_{j} \bigg) \tr(K_T^\dagger K_{N})^*  \bigg \}
%&\quad \textcolor{red}{= 
%2\rre\bigg \{
% \tr \bigg(  K_{T}^\dagger \Big[  \prod_{j'>j} U_{j'} \Big] i\, \delta t\, %\mathcal{H}_k  K_{j} \bigg) \tr(K_T^\dagger K_{N})^*  \bigg \}}
\end{align}
where $\prod$ is understood to produce a time-ordered product. 

% The corresponding backward propagation algorithm generated by automatic differentiation is
% \begin{algorithm}[H]

% \caption{Unitary gate: automatic differentiation}\label{algo:target_unitary_back_prop}

% \begin{algorithmic}[1]
% %\Procedure{Obtaining $\partial C / \partial u_{i}$}{}
% \State $\textit{P} = \tr(K_T^\dagger K_{N})^* K_T^\dagger $
% \For{$j = N $ to 0 }
% \For{all $k$ }
% \State $\partial f_C / \partial u_{k,j} = 2\rre[ \tr ( i \delta t \textit{P} \mathcal{H}_k K_j)] $
% \EndFor
% %\State \textbf{if} $j = 0:$ \textbf{break}
% \State $\textit{P} =  \textit{P} U_{j}$
% \EndFor
% %\EndProcedure
% \end{algorithmic}
% \end{algorithm}

This expression shows that automatic reverse-mode differentiation requires the propagators $K_j$ from every time step. Within TensorFlow, the set of intermediate propagators $\{K_j\}$ is stored in memory during the forward evolution. The resulting memory demand therefore scales as $O(\ell^2 \times N)$. 

\emph{Memory-efficient algorithm.}--- We note that storage of $\{K_j\}$ can be avoided by applying the strategy introduced in the original GRAPE paper \cite{Khaneja2005Optimal}: since the evolution is unitary, one may time-reverse the evolution step by step, and re-calculate the intermediate propagator via $K_{j}=U_{j+1}^\dagger K_{j+1}$. Here, each short-time propagator $U_j$ is re-generated locally in time, using only the control fields at time $t_j$. Such a backwards-propagation algorithm leads to an increase in computation time by roughly a factor of 2 (each $U_j$ is then calculated twice), but has a memory demand of only $O(\ell^2)$ -- which does not scale with $N$, the number of time steps. Thus for large problems the memory efficient algorithm is superior. This memory-efficient algorithm, currently not realized in this implementation, is given by
\begin{algorithm}[H]

\caption{$C_1$ gradient via backwards propagation}\label{algo:target_unitary_memory_eff}

\begin{algorithmic}[1]
%\Procedure{Obtaining $\partial C / \partial u_{j}$}{}
\State $P = \tr(K_T^\dagger K_{N})^* K_T^\dagger $
\State $X = K_{N} $
\For{$j = N $ to 0 }
\For{all $k$ }
\State $\partial C_1 / \partial u_{k,j} = -2\delta t\iim[\tr ({P}\, \mathcal{H}_k\, \textit{X} ) ]$ 
\EndFor
%\State \textbf{if} $j = 0:$ \textbf{break}
\State $X =  U_{j}^\dagger X$
\State $P =  P\, U_{j}$
\EndFor
%\EndProcedure
\State return $\partial C_1 / \partial \mathbf{u}$
\end{algorithmic}
\end{algorithm}
%

%%%%%%%%%%%%%%%%%

\subsection{Gradient for $C_2$: target-state infidelity}
For state preparation or unitaries specified only in a subspace, it is sufficient to optimize the evolution for only a few initial states, rather than for the complete basis. This is achieved by minimizing a cost function based on $C_2(\mathbf{u}) = 1 - |\langle \Psi_T | \Psi_{N}\rangle|^2$, where the realized final state $| \Psi_{N}\rangle$ depends on the control parameters $\mathbf{u}$. Again applying equations \eqref{eq:chain_rule} followed by \eqref{eq:trace_calculus_3} (and using that the trace of a number results in that number), we obtain
\begin{align}\nonumber
&\frac{\partial C_2}{\partial u_{k,j}} = 
%2 \rre \bigg \{
%\tr \bigg[  \langle \Psi_T| ( \prod_{b>j} U_{b} ) i\, \delta t\, \mathcal{H}_k  | \Psi_{j}\rangle  \bigg] \langle \Psi_T | \Psi_{N}\rangle^* 
%\bigg\}
%\\
%&=
-2\, \delta t \iim \bigg [ 
\langle \Psi_T| \Big[  \prod_{j'>j} U_{j'} \Big]  \mathcal{H}_k  | \Psi_{j}\rangle  \langle \Psi_T | \Psi_{N}\rangle^* 
\bigg ]
% \label{eq:target_state_gradient_expression}
\end{align}

\emph{Memory-efficient algorithm.}--- In TensorFlow-based automatic differentiation algorithm here, the intermediate states $\{|\Psi_j\rangle\}$ are stored, leading to a memory requirement of $O(\ell \times N)$, rather than $O(\ell^2 \times N)$ for the full propagators. By using the same backward propagation strategy as above, a more memory-efficient algorithm with memory requirement $O(\ell)$ independent of the time-step number is possible:
\begin{algorithm}[H]

\caption{$C_2$ gradient via backwards propagation}\label{algo:target_state_memory_eff}

\begin{algorithmic}[1]
%\Procedure{Obtaining $\partial C / \partial u_{j}$}{}
\State $P = \langle \Psi_T | \Psi_{N}\rangle^* \langle \Psi_T|  $
\State $X = | \Psi_{N}\rangle $
\For{$j = N $ to 0 }
\For{all $k$ }
\State $\partial C_2 / \partial u_{k,j} = -2\delta t \iim[P\, \mathcal{H}_k\,X ] $
\EndFor
% \State \textbf{if} $j = 0:$ \textbf{break}
\State $X =  U_{j}^\dagger X$
\State $P =  P\, U_{j}$
\EndFor
%\EndProcedure
\State return $\partial C_2 / \partial \mathbf{u}$
\end{algorithmic}
\end{algorithm}

\subsection{Gradient for $C_5$: occupation of forbidden state}
Occupation of a ``forbidden'' state is discouraged by the cost function $C_5=\sum_{j} |\tr( \Psi_F^\dagger \Psi_j)|^2 $. This cost function differs qualitatively from the gate and state infidelity cost functions: the latter are evaluated based on the result at the final time, while forbidden-state occupation involves intermediate states at every time step. Accordingly, the corresponding gradient takes a different form. First, Eq.\ \eqref{eq:chain_rule} is replaced by
\begin{align}
&\frac{\partial}{\partial u_i}c\big(\Psi_0(\mathbf{u}),\Psi_1(\mathbf{u}),\ldots,\Psi_N(\mathbf{u})\big)\nonumber\\ 
&\qquad =  \tr \frac{\partial c}{\partial \Psi_j}\frac{\partial \Psi_j}{\partial u_i} + \text{c.c.}
\label{eq:chain_rule2}
\end{align}
where introduction of the trace of a $c$-number is convenient for direct application of Eq.\ \eqref{eq:trace_calculus_3}. We then obtain
\begin{align}\label{eq:forbidden_state_gradient_expression}
&\frac{\partial C_5}{\partial u_{k,j}} 
\overset{\eqref{eq:chain_rule2}}=
\sum_{J}
\tr \frac{\partial C_5}{\partial \Psi_{J}}\frac{\partial \Psi_{J}}{\partial u_{k,j}} + \text{c.c.}\\\nonumber
&\quad=
\sum_{J\ge j}\sum_{j'}
\tr \bigg[
\frac{\partial [|\tr(\Psi_F^\dag \Psi_{j'})|^2]}{\partial \Psi_{J}}\frac{\partial \Psi_{J}}{\partial u_{k,j}}
\bigg] + \text{c.c.}\\\nonumber
&\quad\overset{\eqref{eq:trace_calculus_3}}=
\sum_{J\ge j} \tr \Big(\Psi_F^\dag\frac{\partial \Psi_{J}}{\partial u_{k,j}}\Big)
\tr(\Psi_F^\dag \Psi_{J})^* + \text{c.c.}
\\\nonumber
%
%&\quad=- i\,\delta t
%\sum_{j_1\ge j}
%\big\langle \Psi_F\big|\big[\prod_{j'>j_1} U_{j'} \big] \mathcal{H}_k \big| %\Psi_{j_1} \big\rangle
%\big\langle \Psi_{j_1}\big|\Psi_F\big\rangle + \text{c.c.}\\\nonumber
%
&\quad=2\,\delta t
\sum_{J\ge j} \iim \bigg[
\big\langle \Psi_F\big|\big[\textstyle\prod_{j'=j+1}^{J} U_{j'} \big] \mathcal{H}_k \big| \Psi_{j} \big\rangle
\big\langle \Psi_{J}\big|\Psi_F\big\rangle \bigg]
\end{align}
% The corresponding backward propagation algorithm generated by automatic differentiation and its memory efficient version would be the following.

The double sum of eq. \eqref{eq:forbidden_state_gradient_expression} makes it appear as though the computation of this gradient would take $O(N^2)$, however after simplification, the relationship between the limits of the sum and product allow it to be calculated in $O(N)$ time. The corresponding backward propagation algorithm then takes the following form:

% \begin{algorithm}[H]

% \caption{Forbidden state: automatic differentiation}\label{algo:forbidden_state_back_prop}

% \begin{algorithmic}[1]
% %\Procedure{Obtaining $\partial C / \partial u_{j}$}{}
% \State $\textit{P} = \langle \Psi_F| \Psi_{N} \rangle^* \langle \Psi_F| $
% \For{$j = N $ to 0 }
% \For{all $k$ }
% \State $\partial f_C / \partial u_{k,j} = - 2 \rre[ i \delta t \textit{P} \mathcal{H}_k | \Psi_j \rangle]$
% \EndFor
% \State $\textit{P} =  \textit{P} U_{j} +  \langle \Psi_F| \Psi_{j-1} \rangle^* \langle \Psi_F| $
% \EndFor
% %\EndProcedure
% \end{algorithmic}
% \end{algorithm}

\begin{algorithm}[H]

\caption{$C_5$ gradient via backwards propagation}\label{algo:forbidden_state_memory_eff}

\begin{algorithmic}[1]
%\Procedure{Obtaining $\partial C / \partial u_{j}$}{}
\State $P = \langle\Psi_{N} | \Psi_F \rangle \langle \Psi_F | $
\State $X = | \Psi_{N} \rangle $
\For{$j = N $ to 0 }
\For{all $k$ }
\State $\partial C_5 / \partial u_{k,j} = 2\,\delta t \iim[  P \mathcal{H}_k X] $
\EndFor
\State $X =  U_{j}^\dagger \textit{X}$
\State $P =  P\, U_{j} + \langle X|\Psi_F\rangle \langle \Psi_F| $
\EndFor
%\EndProcedure
\State return $\partial C_5 / \partial \mathbf{u}$
\end{algorithmic}
\end{algorithm}

This cost function and gradient is also used as the time-optimal award function, using a negative cost to reward rather than penalize the target state at every time step (rather than just at the end). The gradients of cost functions involving only control fields do not involve the time propagation, so we also omit their derivation.

\subsection{Summary}

Algorithms for each cost function along with their computation and memory costs have been presented. The computation time of the algorithms all scale linearly with the number $N$ of time steps. Automatic gradient calculation which requires caching of each step causes memory to scale like $N$, while reducing the run time by a constant factor of 2. By contrast, algorithms which directly exploit the unitary structure of quantum evolution can have memory requirements do not scale with the number of time steps. Hence, it may be worth implementing analytic gradients for very long computations which otherwise would not fit in memory. 

Computing the fidelity and gradient for the whole unitary evolution as in algorithm 1, requires $O(\ell^2)$, whereas state transfer requires $O(\ell)$ memory. It should be noted that full unitary evolution fidelity can also be calculated as $\ell^2$ state transfer computations over a complete basis. This has the memory requirements of state transfer, and the same computation requirements as algorithm 1, though is less efficient by a constant factor. In principle, each state transfer can be performed in parallel and assembled to compute the total cost and gradient. In addition, the Hamiltonians of many physical problems can be represented sparsely allowing a significant speedup in computation as well. For practical problems, the number time steps required may scale with the size of the problem, as more complex quantum gates/algorithms require more time than simple ones. 
%We believe it is preferable to have quick experimentation on new cost functions by automatic differentiation, and then implement memory efficient algorithms for large state optimization.

% ---
% NL : BELOW IS JUST FOR MY OWN REF, WILL REMOVE
%

% \begin{equation}
% \frac{\partial C}{\partial u_{k,j}} = \Bigg[ \sum_{a=j}^{N} \bigg( \frac{\partial f_{C}}{\partial y_{a}} \prod_{b=j+1}^{a} \frac{\partial y_{b}}{\partial y_{b-1}} \bigg) \Bigg] \frac{\partial y_{j}}{\partial u_{k,j}}  
% \label{eq:gradient_expression_gen}
% \end{equation}

% The pseudo-code of backward propagation algorithm to obtain these gradients is

% \begin{algorithm}[H]

% \caption{Backward propagation}\label{algo:back_prop}

% \begin{algorithmic}[1]
% % \Procedure{Obtaining $\partial C / \partial u_{i}$}{}
% \State $\textit{P} = \partial f_{C}/ \partial y_{N} $
% \For{$j = N $ to $0$ }
% \For{all $k$ }
% \State $\partial C / \partial u_{k,j} = \textit{P} \times \partial y_{j}/ \partial u_{k,j} $
% \EndFor
% \State \textbf{if} $j = 0:$ \textbf{break}
% \State $\textit{P} =  \textit{P}\times\partial y_{j}/ \partial y_{j-1} + \partial f_{C}/ \partial y_{j-1}$
% \EndFor
% % \EndProcedure

% \end{algorithmic}
% \end{algorithm}

% \bibliography{nleung91-qoc.bib}


\end{document}